\documentclass[preprint,12pt]{elsarticle}
\usepackage{lineno}

\usepackage{amssymb}
\usepackage{amsmath}
\usepackage{epsfig}
\usepackage{graphicx}
\usepackage{epstopdf}
\usepackage[figuresright]{rotating}
\usepackage{multirow}
\usepackage{url}

\journal{NIM}

\begin{document}
\begin{frontmatter}





\title{Radiation Testing of Electronics for the CMS Endcap Muon System}
\author[osu]{B.~Bylsma}
\author[tamu]{D.~Cady}
\author[tamu]{A.~Celik}
\author[osu]{L.~S.~Durkin}
\author[tamu]{J.~Gilmore}
\author[neu]{J.~Haley}
\author[tamu]{V.~Khotilovich}
\author[tamu]{S.~Lakdawala}
\author[rice]{J.~Liu}
\author[rice]{M.~Matveev}
\author[rice]{B.~P.~Padley}
\author[rice]{J.~Roberts}
\author[tamu]{J.~Roe}
\author[tamu]{A.~Safonov}
\author[tamu]{I.~Suarez}
\author[neu]{D.~Wood}
\author[tamu]{I.~Zawisza}

\address[tamu]{Texas A\&M University, College Station, TX 77843}
\address[neu]{Northeastern University}
\address[rice]{Rice University}
\address[osu]{Ohio State University}

\begin{abstract}
The electronics used in the data readout and triggering system for the Compact Muon Solenoid (CMS) experiment at the Large Hadron Collider (LHC) particle accelerator at CERN are exposed to high radiation levels. This radiation can cause permanent damage to the electronic circuitry, as well as temporary effects such as data corruption induced by Single Event Upsets. Once the High Luminosity LHC (HL-LHC) accelerator upgrades are completed it will have five times higher instantaneous luminosity than LHC, allowing for detection of rare physics processes, new particles and interactions. Tests have been performed to determine the effects of radiation on the electronic components to be used for the Endcap Muon electronics project currently being designed for installation in the CMS experiment in 2013. During these tests the digital components on the test boards were operating with active data readout while being irradiated with 55~MeV protons. In reactor tests, components were exposed to 30 years equivalent levels of neutron radiation expected at the HL-LHC. The highest total ionizing dose (TID) for the muon system is expected at the inner-most portion of the CMS detector, with 8900~rad over ten years. Our results show that Commercial Off-The-Shelf (COTS) components selected for the new electronics will operate reliably in the CMS radiation environment.
\end{abstract}
\end{frontmatter}
\section{Introduction}
The Compact Muon Solenoid (CMS) experiment~\cite{cms2008} is one of the two large general-purpose experiments at the Large Hadron Collider (LHC). The purpose of this experiment is to understand the mechanism of electroweak symmetry breaking responsible for generating masses of particles and search for evidence of new physics, such as supersymmetry and extra dimensions. The CMS detector is one of the most advanced particle detectors ever built, and is comprised of several distinct systems. These include the silicon tracker, electromagnetic calorimeter, hadronic calorimeter, super-conducting solenoid and muon chambers.  Working together, these systems identify particles as well as measure their energy and momenta.

In order to maximize the amount of analyzable data, the beam intensity of LHC will be increased, leading to five times higher data rate in CMS.  With this forthcoming High Luminosity LHC (HL-LHC) accelerator upgrade, the Cathode Strip Chamber (CSC) muon detectors~\cite{cms-muon} in the CMS endcap will require new electronics to handle the increased rate while maintaining high data collection efficiency.

The increased rate of collisions will also raise the radiation level experienced by the CMS data readout and trigger electronics. Such exposure results in both cumulative effects and Single Event Upsets (SEUs) that degrade the performance of silicon circuits. The electronic components used in building the new CSC data readout and trigger system must be designed to reliably handle the high data rate while operating in a high radiation environment.

Therefore it is necessary to determine which commercial electronic components are safe for high luminosity operation in the CMS environment, and this intial study addresses the most critical concerns in a series of tests that were carried out in the Radiation Effects Facility at Texas A\&M University (TAMU) Cyclotron Institute and at the TAMU Nuclear Science Center (NSC). The K500 cyclotron at TAMU provides a proton beam with 55~MeV energy, while the reactor at the NSC provides neutrons with energy up to a few MeV. A similar study was made elsewhere~\cite{osu-paper} in 2001 to evaluate the components currently used by the CSC system in the first phase of LHC running.

The remainder of this paper is organized as follows: In section 2, we describe the components of the new electronics for the CSC system. In section 3, we review the expected CMS radiation levels. The radiation testing setups for digital and non-digital components are discussed in section 4. Results and conclusions are presented in sections 5 and 6, respectively. 

\section {Design Considerations for New Cathode Strip Chamber Electronics}
The CSCs are a critical component of CMS that are used to identify and measure momentum of muons passing through the detector in the most forward region. The electronics used to collect CSC data must be capable of handling the high data rate and withstand the radiation environment in the CMS endcap. New electronics are being designed to replace the existing electronics for the innermost region of the CSC system, specifically the 36 ``ME1/1'' chambers on each endcap.  The design includes improvements in data transmission and logic capability to ensure efficient performance.

There are three primary components of the ME1/1 CSC electronics being replaced; Cathode Front End Boards (CFEB), Trigger Motherboards (TMB) and Data Motherboards (DMB). These last two types are custom VME boards~\cite{vmebus} that operate in an electronics crate a few meters away from the CSCs to which they are connected. All of these boards are constructed using Commercial Off The Shelf (COTS) electronic components such as FPGAs, voltage regulators etc. The CFEBs provide digitized strip data from a CSC; the TMB receives a coarse sampling of this data to perform fast pattern matching and generates a signal to trigger the readout of the detector when it determines that a muon track has passed through the CSC. When that trigger signal is received by the CFEBs, the fully digitized muon track data is sent to the DMB, and then on to the CMS central data acquisition and storage system. The continued high level of efficiency in the system depends upon implementation of new algorithms in these boards, which requires improvements such as new, powerful FPGAs and new optical data links in order to handle the increased rate of data.

Maintaining high efficiency in the system also requires increased segmentation of the chamber electronics. In the current configuration of CSCs, each of the innermost (and the highest rate) muon chambers carry five CFEBs, which are read out via copper cables. In the new design there will be seven new high-performance ``Digital CFEBs'' per chamber, which will reduce the duty factor for each board, eliminate dead-time, and improve processing efficiency.  Geometrical constraints make large copper cables unsuitable for this expansion to seven CFEBs, and optical links are the only viable readout alternative. Furthermore, improved logic capability is required for all three board types to handle the additional data channels; this is addressed by utilizing modern FPGAs with increased logic and memory resources.

Radiation effects must be considered in the design process as these can cause electronics malfunctions. The most significant effects to consider in the CMS endcap fall into two general categories: Single event effects and cumulative effects. Single event effects can occur as a Single Event Upset (SEU) or as a latchup condition. SEUs are transient disruptions in digital circuits, causing no permanent damage; they are easily recoverable by simply writing or resetting the logic registers. However, a high rate of such errors can cause unacceptable levels of dead time and reduce efficiency of the system, so it is important to quantify the rate of SEUs expected for each component.

Unlike an SEU, a latchup condition can cause permanent chip failure.  Latchup is generally characterized by a large increase in current drawn by a chip, and the subsequent overheating can burn out elements in the silicon. Occurences such as this would require a time consuming intervention to replace the damaged parts, so it is important to design the CSC electronics with components that are not susceptible to latchup in the CMS endcap environment.

Cumulative effects are permanent incremental damage attributed to total ionizing dose (TID) or displacement damage, which degrades the performance of silicon circuits over time~\cite{cots-faccio}. Several components were tested in this study with different manufacturing technologies, and there are inherent differences in their susceptibility to cumulative effects. This study makes no attempt to distinguish between TID and displacement damage failures, but the final results on performance will be used to determine which components are likely to survive the radiation environment in the CMS endcap. Any COTS components found to be unsatisfactory for any reason will not be used in the new design.

Single event effects are caused by hadronic interactions in the silicon that change the logical state of a digital circuit\footnote{We distinguish digital parts by their fundamental function, in that they primarily handle binary information data, although in some cases they may be mixed internally with analog elements.}. In the CMS endcap environment, at least 90\% of these are caused by neutrons with energy above 20~MeV~\cite{seu-sim}. The authors in \cite{seu-sim} note that for energy above 20~MeV, the upset suscepibility in silicon is equivalent for protons and neutrons. Thus, for digital components, both TID/displacement damage as well as SEU and latchup susceptibility can be evaluated at the same time using a cyclotron beam that provides protons with energy greater than 20~MeV.  However, for the non-digital components in the CSC system the biggest concern is TID/displacement damage, so exposure to neutrons with energy around 1~MeV from a nuclear reactor is largely sufficient as a measure of their radiation tolerance.

The measurements made in this study show the impact of SEUs in different components; e.g. most errors in the optical receivers are transient, affecting only a single data word with little overall significance, but a complex component like an FPGA may require a reset for SEU recovery, and this causes dead time.  Information about the SEUs rates and the impact of any recovery action are combined for use as input in determining where mitigation is required in the system. These factors are discussed individually for each part later in the description of test results.

\begin{figure}[htb]
\centering
\includegraphics[width=8cm,height=7cm] {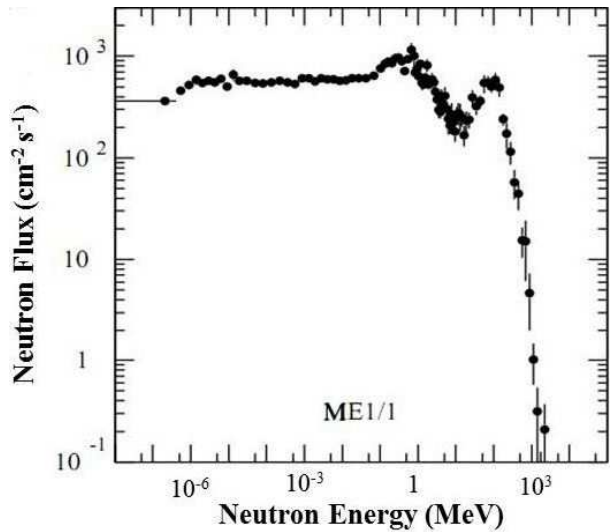} 
\caption{Energy spectrum of neutron exposure for the innermost chambers of the CSC system, based on simulations from the first phase of LHC operation~\cite{cots-huhtinen}.\label{fig:neutron-spectrum}}
\end{figure}

Survivability is also important, and the requirements for this are discussed further in the next section. However, TID and displacement damage results for digital components are not included here, as the flux of the TAMU cyclotron was not sufficient to accumulate significant dose\footnote{Full TID/displacement damage testing for digital components is planned for a future study.}.  Similarly, the non-digital parts were not tested for SEU sensitivity due to the large sample of initial candidates; however, the best candidates determined in this study will be subject to additional power-on tests in the future to verify their tolerance to SEUs as well as cumulative effects.

\section {CSC Radiation Level In CMS}
\begin{figure}[htb]
\centering
\includegraphics[width=8cm,height=7cm] {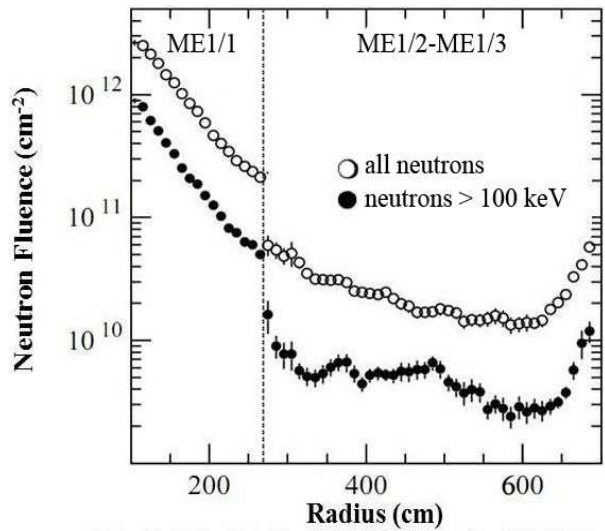} 
\caption{Expected neutron fluence in the Station 1 region of the CMS endcap for the first phase of LHC operation, shown as a function of the distance from the beam line~\cite{cots-huhtinen}.\label{fig:neutron-fluence}}
\end{figure}

ME1/1 CSCs are the innermost chambers within CMS and are exposed to the highest rate of radiation damage, largely from neutrons. Figure~\ref{fig:neutron-spectrum} shows the simulated energy spectrum of neutrons crossing the ME1/1 region; figure~\ref{fig:neutron-fluence} shows how the neutron fluence in the endcap region is expected to vary with distance from the beam line. In the area of highest neutron flux, the total fluence for neutrons with energy E$>$100~keV is about $6\times 10^{11}$~cm$^{-2}$ \cite{cots-huhtinen}, and the TID is 1780~rad over ten years of normal LHC operation. However, for HL-LHC these rates are expected to be five times higher, as summarized in table~\ref{tab:exposure-summary}. These data form the baseline for determining the radiation tolerance of the CSC electronics.

\begin{table*}[ht]
\centering
 \begin {tabular}{c|c}
\multicolumn{2}{c}{Ten Year HL-LHC Exposure in ME1/1} \\ \cline{1-2}  
 20~MeV neutron fluence & $ 2.7 \times 10^{11}$ n/cm$^2$ \\
 1~MeV neutron fluence & $ 3.0 \times 10^{12}$ n/cm$^2$  \\
 total ionizing dose & $8.9$ krad \\
\end{tabular}
\caption{Summary of expected neutron exposure in the ME1/1 region of the CMS endcap for ten years of HL-LHC operation. \label{tab:exposure-summary}}
\end {table*}

\section{ Radiation Testing Setup}
Radiation tests were performed separately for digital and non-digital components at the TAMU cyclotron and the NSC to determine their radiation tolerance. The testing facilities, equipment and setup are discussed for both digital and non-digital components under sections 4.1 and 4.2 respectively.  A special distinction is made for digital components as they have a significant susceptibility to SEU errors in data handling and control logic.

\subsection {Digital Components}
\begin{figure}[htb]
\centering
\includegraphics[width=11cm,height=7cm] {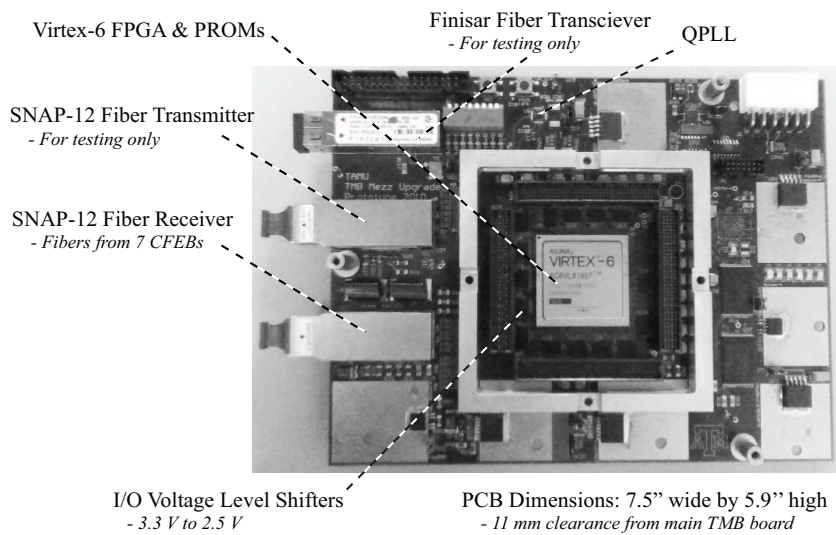}
\caption{Top-side view of a TMB mezzanine test board, as used in SEU tests for digital components.\label{fig:tmb-mezz}}
\end{figure}

The proposed CSC electronics replacement introduces several new digital devices to the system that require testing. These include the Xilinx Virtex 6 FPGAs, the Finisar duplex opto-coupler, Reflex Photonics Snap12 simplex receiver and transmitter opto-couplers, and the TI SN74CB3T bus exchange level shifter. A printed circuit board containing these components was developed and assembled (fig.~\ref{fig:tmb-mezz}) as a test bed for the measurements of radiation tolerance. During the board design, special attention was given to the location of the parts on the board, allowing sufficient space for a collimated beam to irradiate each component one-by-one with no overlap.

In this study the digital components were tested for SEU sensitivity\footnote{These digital parts are primarily CMOS components that should be less sensitive to damage at the moderate levels of radiation exposure expected at HL-LHC. However, the beam flux available in this study was not sufficient to prove their survivability, so the TID/displacement damage tolerance must be fully evaluated with additional testing.}.  SEUs are caused primarily by neutrons with energy above 20~MeV, and the fluence of such neutrons in the ME1/1 region of the CMS endcap is expected to be $2.7 \times 10^{11}$~cm$^{-2}$ over ten years of HL-LHC running.  However, for the purpose of this test it is only necessary to observe enough SEUs to make a statistically signifiacnt measurement of the SEU cross sections. To accomplish this, the total number of observed operational errors for each component were counted (regardless of mechanism) while they were exposed in a controlled proton beam. These functional SEU cross section measurements are then used to evaluate the SEU sensitivity for each part and determine the level of error mitigation that may be required.

The TAMU cyclotron provided a 55~MeV proton beam with a flux as high as $3 \times 10^7$~cm$^{-2}s^{-1}$, and uniform across the 1.5~inch diameter beam spot. The flux measurements were made using system of scintillation counters and a Faraday cup, providing a precision better than 1\%. A movable platform was used to support the TMB mezzanine test boards in perpendicular orientation with respect to the beam line, and programmable stops were set for the location of each device under test (DUT) to ensure alignment in the beam. Two samples were tested for each of the digital COTS components planned for use in the new CSC electronics.

System checks were performed just before irradiating each board to verify that the power was stable and the SEU error monitoring system (described below) was operational. Beam intensity for each DUT began with a base level flux of $1 \times 10^{7}$~cm$^{-2}s^{-1}$. The SEU rate was monitored during the first few minutes of each DUT exposure, and the beam flux was adjusted as needed to attain a suitable rate of SEUs, typically up to ten per minute. Each DUT was exposed for 45 to 90 minutes in the beam.  Power supply currents were monitored for signs of latchup during exposure.

A sophisticated error monitoring system was developed, using custom firmware and software, in order to detect the occurrence of SEUs and to reliably identify the exact element that failed among the several different digital components mounted on the TMB mezzanine test board.  The specific tests employed in each case are described individually in the ``Results'' section, but every error event detected was automatically verified in real-time by performing multiple readbacks from the hardware to collect information associated with the SEU event, and checking it for consistency in the software.  For example, errors in the Block RAMs (programmed as ROMs) usually involved one or more bit flips, and in these cases the memory was read back again to confirm that there was bad data in the memory, and not a transmission error.  All the information about each SEU event was saved in a log file for later review, including a time stamp, the error data, what the data should have been, and which test condition was activated by the SEU.

\subsection {Non-Digital Components}
\begin{figure*}[htb]
\centering
\includegraphics[width=7cm,height=9.2cm,angle=270] {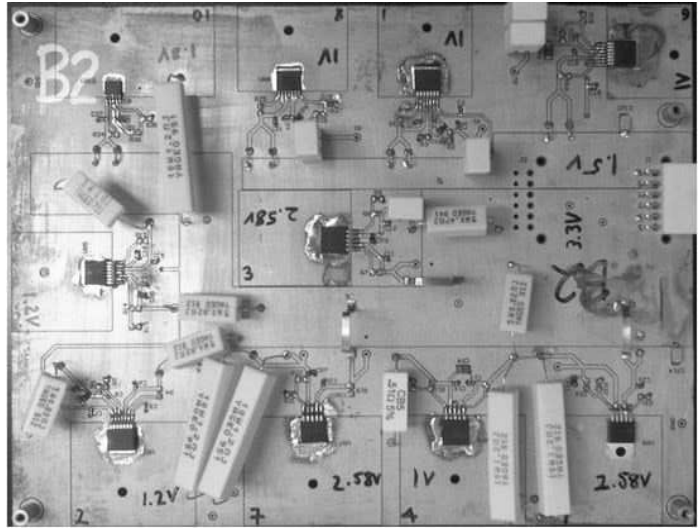}
\caption{One of the voltage regulator test boards that was exposed to 1~MeV neutrons in the TAMU NSC reactor.\label{fig:reg-test}}
\end{figure*}

There are several non-digital components required in the new CSC electronics, such as voltage regulators and diodes which are necessary for providing power to the digital components. Custom designed voltage regulator test boards were assembled (fig.~\ref{fig:reg-test}) for testing the performance of multiple samples for each component.

Based on the radiation exposure calculations for HL-LHC, the electronics mounted on ME1/1 chambers must withstand 1~MeV neutron fluence of $3 \times10^{12}$~cm$^{-2}$ and about 9~krad total dose over ten years. To account for the uncertainties in these calculations, a safety factor of three times the expected levels was imposed, and these radiation tests extend beyond 30~krad.

The Texas A\&M Nuclear Science Center supports a 1 Megawatt TRIGA reactor that serves as an ideal environment to determine those parts which are suitable for use in the CSC electronics project. The voltage regulator test boards received an initial exposure for one hour at 6 kilowatt reactor power, with a 1~MeV neutron flux of $9.9 \times 10^{8}$~cm$^{-2}s^{-1}$; a few weeks later they were exposed for an additional two hours at same reactor power. The accumulated radiation exposure levels correspond to 10 and 30 HL-LHC years respectively (approximately 10~krad and 30~krad), and the total neutron fluence was $10.7 \times 10^{12}$~cm$^{-2}$.

\section{Results}

\begin{figure}[htb]
\centering
\includegraphics[width=10cm,height=5cm] {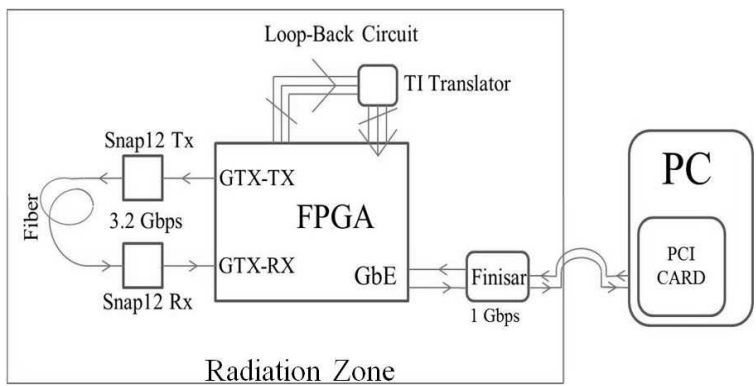}
\caption{Diagram of the test circuits used on the TMB mezzanine test boards, indicating the five digital devices tested in the TAMU cyclotron.\label{fig:digi-test}}
\end{figure}

This section provides a description for each type of component tested and explains the procedures used to carry out the tests in each case; an overview of the test configuration used for the digital components is shown in figure~\ref{fig:digi-test}.  For the non-digital parts, each was tested with dedicated load resistors tuned to emulate the realistic conditions for normal CMS operation.
Results are presented for SEU susceptibility and radiation survivability, based on measurements such as SEU counts, voltage levels, total fluence and the dose during each exposure. 

It was noted that a few SEUs were observed in the FPGA during exposure of other components, likely caused by backscatter from the beam.  These kind of SEUs were clearly distinguished in the error monitoring system and they were allocated to the appropriate SEU category. Furthermore, the rate of these was very low, causing less than 1\% of all SEUs, so the overall impact on the SEU test results is negligible.

Power supply currents were monitored during exposure and no changes were observed, which is evidence that no latchup occurred in this study.

\subsection {Snap12 Receiver/Transmitter: Reflex Photonics SN-R12-C01001/T12-C01001}
The Snap12 transmitter and receiver are used to transfer data through 12 independent optical channels. For the model tested, each channel is capable of operating at 6.25~Gb/s through 300~m of fiber optic cable at a wavelength of 850 nm. The transmitter converts 12 electrical signals into fiber optic signals, while the receiver performs reversed conversion. In most cases, the new CSC electronics will not utilize all 12 of these links; only 6 were tested on each board in this test.

During the test, the FPGA was programmed to use a Pseudo-Random Bit Generator (PRBG) to create randomized data, and transmit the data at 3.2~Gb/s through the Snap12 devices. The Snap12 fibers were physically looped-back from the output (transmit) port to an input (receive) port with a 15~m long MTP fiber ribbon cable.  The FPGA received the data and checked for differences against what was sent; six such fiber circuits were operated in parallel for this test. Any discrepancy between transmitted and received data was identified as an SEU, and a count was made of such occurrences. These counts were read out from the FPGA via a gigabit Ethernet (GbE) link to a PC and monitored by software at 1 sec intervals. Software recorded time stamp information associated with each incremental error.

Test results for Snap12 receiver and transmitter are shown in table~\ref{tab:seu-results}.  The typical SEU failure mode was due to bit errors in a single data word.  In HL-LHC these transient errors will occur at a low rate, with only about one per week on each receiver channel and almost zero errors from the transmitters, so no mitigation is needed for the Snap12 components.

\subsection {Optical Transceiver: Finisar FTLF8524E2GNL}
Optical transceivers are used to build high-speed data links over multi-mode fiber optics. The Finisar component selected for the new electronics is bidirectional and can operate at rates up to 4.25~Gb/s.

For radiation testing, the FPGA was programmed to send GbE packets carrying randomized data through the Finisar. The fiber was physically connected to a commercial PCI gigabit Ethernet card in the test control PC, and the data packet integrity was tested using a cyclic redundancy check~\cite{crc-book}. After each packet was checked, a new packet was transmitted. Any discrepancy between transmitted and received data packets was considered an SEU, and such error instances were counted.

In cases of persistent error the test operator issued a system reset; if not resolved by a reset, the operator turned off the power supply for five seconds, then resumed testing.  A power cycle was required a total of eight times due to the loss of the data link on the Tx channel.  No change in current was observed, so an SEU in the transmit enable circuit seems the likely cause rather than latchup. The cross section of this occurence is $5.9 \times 10^{-11}$~cm$^2$ which corresponds to about 1.5 times per year for each Finisar component during real operations at CMS.  While this is rate is acceptable for use in the new system, it will be investigated further in a future study to verify the cause of the problem and to establish a reliable mechanism for mitigation.

The test result summary for the Finisar is shown in table~\ref{tab:seu-results}.  However, deeper analysis of the errors shows that the rate of SEU induced bit errors scales with the rate of packets transmitted on the GbE link.  Rescaling this SEU crosss section for the worst-case data load expected at HL-LHC increases it to $4.5 \times 10^{-8}$~cm$^2$ which corresponds to about seven errors per day on each link during real operation at CMS. This is less than the typical industry standard for bit errors on gigabit links of one error per 10~trillion bits, and such low rate of transient bit errors is acceptable in CMS with no need for mitigation.

\subsection {Bus Exchange Level Shifter: Texas Instruments SN74CB3T16212}
While newer electronics used for the CSC project rely on 2.5~V technology, they must also communicate with some older 3.3~V components in the system. This is a concern for the Virtex 6 FPGA which is limited to 2.5~V signal levels. To keep the chip safe, the voltage level of the incoming signals to the FPGA must be reduced to 2.5~V using the TI level shifter.

To test these chips for SEU susceptibility, the FPGA was programmed to transmit randomized data patterns through the DUT. A set of outgoing FPGA data lines were electrically looped back into the level shifter input port using a customized circuit board. The data then passed from the level shifter to the FPGA, where it was checked for differences against what was sent.  As each received word was checked, a new word was transmitted, at 80 MHz clock frequency.  As above, any discrepancy between transmitted and received data words was identified as an SEU; the FPGA counted such instances in an internal register, which was accessed at 1 sec intervals via GbE link to a PC. Software recorded the time stamp and other information associated with each error. Since no errors were observed, the test results shown in table~\ref{tab:seu-results} are the upper limits on the SEU cross section at 90~\% confidence level.

\subsection {Xilinx Virtex-6 FPGA: XC6V195T-2FFG1156CES}
Virtex 6 is one of the most advanced FPGAs available, containing many specialized built-in-silicon modules, and it requires several voltage supplies to operate, ranging from 1.0~V to 2.5~V. Having more logic resources than previous models, this FPGA is capable of processing data more efficiently. It also has dedicated memory blocks and multi-gigabit transceiver ``GTX'' blocks. Using the TAMU cyclotron, two Virtex 6 chips were tested and the results for the embedded silicon modules are indicated in table~\ref{tab:seu-results}. The FPGA firmware design for these tests utilized 11 out of 20 GTX blocks, 74$\%$ of the Block RAMs, and 38$\%$ of the FPGA Configurable Logic Blocks (CLBs).  In order to measure the maximum level of FPGA sensitivity to SEUs, no SEU mitigation logic was implemented during the tests, and every error was counted as an SEU, regardless of the mechanism causing it\footnote{There is no distinction made between SEUs in the logic fabric and the configuration memory of the FPGA; any error detected in FPGA function is counted as an SEU.}.

The FPGA CLBs and Block RAM memories were configured as ROMs and preloaded with known sets of randomized data. These ROMs were repeatedly read out to a PC using a GbE link, and data checks were performed in software.  Discrepancies found in the data were identified as SEUs; the software saved the information associated with each error in a log file (including time stamp, the identity of the wrong bits and the raw data itself) and notified the test operator that a system reset was required to recover the corrupted ROM.  Resets were controlled manually by push button. 

The FPGA GTX tests occurred in parallel with FPGA logic and memory testing\footnote{These are separate test modules in the firmware operating simultaneously in the FPGA, and the errors for each are tracked individually.} during FPGA proton beam exposure. Custom error monitoring tools were used to independently track the different types of errors. Test results for all of the Virtex-6 FPGA components are shown in table~\ref{tab:seu-results}.

Analysis of the recorded SEU events showed that in most cases, the observed Block RAM errors differed from the original preloaded patterns by a single bit.  In cases with differences in more that one bit, the corruption could be attributed to the inversion of a single bit in the logic fabric of the FPGA used for Block RAM control.

\begin{table*}[h]
\tiny
 \begin {tabular}{|l|c|c|c|c|c|c|}
\hline
\multirow{2}{*}{Component tested} & \multicolumn{2}{|c|}{Board 1} & \multicolumn{2}{|c|}{Board 2} & Average & Expected SEU \\ \cline{2-6}
   & SEU & $\sigma( \times 10^{-11}$cm$^2)$ & SEU & $\sigma( \times 10^{-11}$cm$^2)$ & $\sigma_{avg}( \times 10^{-11}$cm$^2)$ & Rate at CMS \\ \hline
Finisar & 6 & $8\pm3 $  & 8  & $13\pm5$  & $10\pm3$  & 3/year/chip \\
\hline
Snap12 Tx & 6 & $9\pm3$  & 3  & $6\pm3$  & $7\pm2$ & $<1$/year/link \\
\hline
Snap12 Rx& 278 & $850\pm50$  & 311  & $790\pm50$  & $820\pm30$ & 4/month/link \\
\hline
TI Level Shifter & 0 & $<4.0$  & 0 & $<3.0$  & $<1.7$ & 0 \\
\hline
FPGA-GTX & 52 & $80\pm10$  & 39  & $70\pm10$  & $76\pm8$ & 3/year/link \\
\hline
FPGA-CLB & 40 & $3800\pm600$  & 22  & $3600\pm800$  & $3700\pm500$ & 6/day/chip \\
\hline
FPGA-Bram & 61 & $5800\pm700$  & 34 & $5500\pm900$  & $5700\pm600$ & 9/day/chip \\
\hline
\end{tabular}
\caption{Number of SEUs observed for each component during proton irradiation and the resulting cross sections.  The last column shows the expected rate of SEUs during HL-LHC operation for each device if no mitigation is implemented. \label{tab:seu-results}}
\end {table*}

The measured GTX error rate is relatively low, with only three transient errors expected per year on each link at Hl-LHC, so no dedicated mitigation is required for these elements of the FPGA.  However, the Block RAM and CLB logic error rate is significant, with several errors expected per chip every day. Block RAM errors can be mitigated using the error detection and correction features embedded in the FPGA. CLB logic errors must be mitigated by implementing triple module redundancy, where voting is used to correct bit errors.  The effectiveness of these methods is to be tested in a future study.

\subsection {Voltage Regulators}
Eleven different voltage regulator models from five different manufacturers were mounted on custom-designed voltage regulator test boards (fig.~\ref{fig:reg-test}). The boards were placed in the large-sample irradiation chamber, and they were not powered during exposure. The voltage, current and temperature performance of each circuit was tested before and after each exposure in order to determine their radiation tolerance.  At least two samples of each regulator were tested, and consistent pattern was observed in the results.

\begin{figure*}[htb]
\centering
\includegraphics[width=11cm,height=10cm] {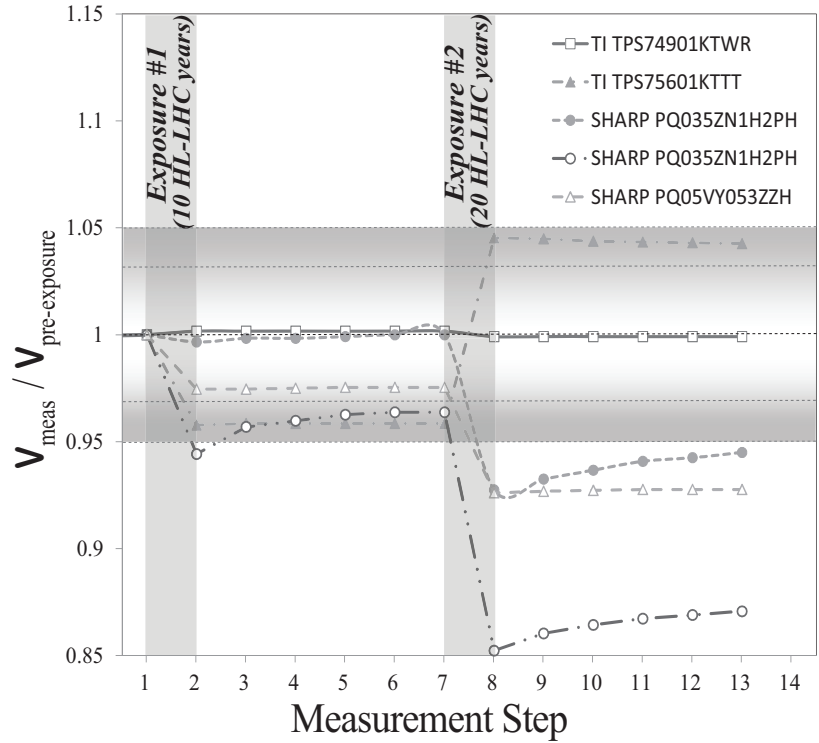}
\caption{Examples of voltage regulator output stability after each radiation exposure, showing the change in voltage output for regulators that survived the full exposure level.  After each exposure the output voltages were measured under power at 30 minute intervals. Voltage regulator deviations greater than 5\% are not acceptable for CSC electronics.\label{fig:reg-plots}}
\end{figure*}

In figure~\ref{fig:reg-plots}, a representative sample of results from five different regulators is shown. The vertical axis represents the ratio of the measured voltage (after exposure) to the pre-exposure voltage; the horizontal axis shows the measurements taken in 30-minute time intervals after each radiation exposure. Observing a constant result with value $\approx 1$ over time indicates that the regulator is a stable, reliable choice for our design. As seen from the figure, several regulators break down after exposure and fail to regulate; some of the samples perform well after the first 10~krad exposure, and then fail after the final 20~krad exposure. Some chips show a small degree of recovery over time, but they remain outside of the acceptable performance range of $\pm 5 \%$ voltage variation. 

Three of the six Max8557ETE regulators showed 0~V output after the second exposure, while the other three show approximately 3\% decrease in voltage. This is the only observed case of complete failure for a regulator due to radiation. Several other regulators showed more than 5\% change in voltage, and regardless of the mechanism, these are all unacceptable for use in CSC electronics.

The following regulator models demonstrated performance degradation to varying degrees and do not show the tolerance required for use in CMS:

\begin{itemize}
  \item[{\huge $\cdot$}] Maxim 8557ETE (p);
  \item[{\huge $\cdot$}] Sharp PQ05VY053ZZH, PQ035ZN1H2PH, and PQ070XZ02ZPH;
  \item[{\huge $\cdot$}] TI TPS75601KTT (p) and TPS75901 (p).
\end{itemize}
where (p) indicates parts manufactured with p-channel MOSFETs.

About half of the tested regulator models showed no significant change in performance, and these voltage regulators were determined to be suitable for use in the new CSC electronics\footnote{The initial sample size of regulators was too broad for efficient evaluation of SEU tolerance at the TAMU cyclotron, but this will be addressed for the surviving candidates in a future study.}:
\begin{itemize}
  \item[{\huge $\cdot$}] National Semi LP38501-ADJ (n) and LP38853S-ADJ (n);
  \item[{\huge $\cdot$}] TI TPS74901KTWR (n);
  \item[{\huge $\cdot$}] MIC69502WR and MIC49500WU.
\end{itemize}
where (n) indicates parts manufactured with n-channel MOSFETs.

This reduced set of suitable candidates will undergo further testing in the future to verify their tolerance to SEUs and cumulative effects under power.

\subsection {ST Micro and ON Semi Diodes: 1N5819}
The CSC low voltage supply system utilizes 1N5819 diodes for power supply protection, and these need testing just as the other silicon components. Ten samples of these diodes from two different manufacturers were tested in conjunction with the voltage regulator tests. Performance testing was done before and after each exposure. The performance validation circuit used a 7.67~V DC supply with a series resistor of 7.5~$\Omega$ that was connected through the diode to ground.  Each diode test consisted of two trials: one with forward bias and another with reverse bias. A small voltage drop across each diode was observed in the forward trials, as shown in table~\ref{tab:diode-results}; in the reverse direction there was zero current through the circuit. All diodes showed consistent diode properties after neutron irradiation, with essentially no degradation from the radiation.

\begin{table*}[h]
\centering
 \begin {tabular}{cccc}
 &PRE RAD. &POST RAD. I &POST RAD. II\\

Manufacturer&  $V_{fwd}\; (V)$   &$V_{fwd}\; (V)$  &$V_{fwd}\; (V)$   \\
\hline
ST Micro& 0.413&0.421& 0.413\\

ON Semi& 0.391&0.395&0.388\\

\end{tabular}
\caption{Average measurements of the forward diode voltages before and after each exposure.\label{tab:diode-results}}
\end {table*}

\section {Conclusion}
Several COTS components planned for use in the CSC electronics project were tested at the TAMU Cyclotron Institute and the TAMU Nuclear Science Center reactor. Radiation tolerant power handling components were identified that are likely to be suitable for use in the CMS Endcap Muon system.  Some unsuitable components were also identified during these tests.

SEU susceptibility was measured for several digital components.  The results show that most of these parts will not need SEU mitigation in the CMS endcap environment during HL-LHC operation; however, the need for mitigation was identified for the logic and memory elements in the Xilinx Virtex-6 FPGA.

\section {Acknowledgements}
The authors are grateful to the U.S. CMS Operations Program and U.S Department of Energy in supporting this work. The efforts of the staff at the TAMU Cyclotron Institute and the TAMU Nuclear Science Center are also greatly appreciated.\\
\newpage

\def\NCA{Nuovo Cimento}
\def\NIM{Nucl. Instrum. Methods}
\def\NIMA{{Nucl. Instrum. Methods} A}
\def\NP{Nucl. Phys.}
\def\NPB{{Nucl. Phys.} B}
\def\PLB{{Phys. Lett.}  B}
\def\PRL{Phys. Rev. Lett.}
\def\RPP{Rep. Prog. Phys.}
\def\PRD{{Phys. Rev.} D}
\def\PR{Phys. Rep.}
\def\PRP{Prog. Theor. Phys.}
\def\ZPC{{Z. Phys.} C}
\def\MPL{{Mod. Phys. Lett.} A}
\def\EPJC{{Eur. Phys. J.} C}
\def\CPC{Comput. Phys. Commun.}



\end{document}